\documentclass[pre,floatfix,amsmath,twocolumn,preprintnumbers]{revtex4}

\usepackage{graphicx}

\newcommand{\oscoeff}{\widehat{p}}

\newcommand{\rand}{{\operatorname{rand}}}

\newcommand{\romev}{{\operatorname{ev}}}

\newcommand{\romB}{{\operatorname{B}}}
\newcommand{\romC}{{\operatorname{C}}}
\newcommand{\romr}{{\operatorname{r}}}

\newcommand{\eg}{$e$.$\,g$.}
\newcommand{\ie}{$i$.$\,e$.}
\newcommand{\etal}{{$et.\,al$}}

\begin{document}


\title{Attraction and ionic correlations between charged stiff polyelectrolytes}

\author{Markus Deserno}
\affiliation{Department of Chemistry and Biochemistry, UCLA, USA}

\author{Axel Arnold}
\author{Christian Holm}
\affiliation{Max-Planck-Institute for Polymer Research, Ackermannweg 10, 55128 Mainz, Germany}

\date{June 10, 2002}

\begin{abstract}
  We use Molecular Dynamics simulations to study attractive
  interactions and the underlying ionic correlations between parallel
  like-charged rods in the absence of additional salt.  For a generic
  bulk system of rods we identify a reduction of short range
  repulsions as the origin of a negative osmotic coefficient.  The
  counterions show signs of a weak three-dimensional order in the
  attractive regime only once the rod-imposed charge-inhomogeneities
  are divided out.  We also treat the case of attraction between a
  single pair of rods for a few selected line charge densities and rod
  radii.  Measurements of the individual contributions to the force
  between close rods are studied as a function of Bjerrum length. We
  find that even though the total force is always attractive at
  sufficiently high Bjerrum length, the electrostatic contribution can
  ultimately become repulsive.  We also measure azimuthal and
  longitudinal correlation functions to answer the question how
  condensed ions are distributed with respect to each other and to the
  neighboring rod.  For instance, we show that the prevalent image of
  mutually interlocked ions is qualitatively correct, even though
  modifications due to thermal fluctuations are usually strong.
\end{abstract}

\maketitle


\section{Introduction}

The interaction of polyelectrolytes or charged colloids in polar
solvent depends sensitively on the structure of the electrical double
layer surrounding the macroion. This double layer consists in the
simplest case of small counterions of opposite charge.  The linearized
mean-field treatment of this layer lies at the heart of DLVO theory
\cite{DeLa41, VeOv48}, one of the most influential and still very
important descriptions of these systems.

Today it is well established that sufficiently strong electrostatic
interactions entail phenomena qualitatively beyond the mean-field
level.  The possibility of attractive interactions between
like-charged cylindrical macroions due to correlated ion fluctuation
has been noted as early as 1968 by Oosawa \cite{Oos68, Oos71}, and
Patey showed in 1980 (using an integral equation approach and the HNC
closure) that two like-charged spheres will ultimately attract
\cite{Pat80}.  For various reasons these results were initially
viewed with some scepticism, but attractive interactions and other
non-mean-field phenomena (like overcharging) have soon after been
confirmed by a large number of studies, based for instance on computer
simulations \cite{BrVl82, GuJo84, SvJo84, GuNi86, NiGu91, VaIv91,
LyNo97, GrMa97, LyTa98, AlDa98, GrBe98, Ste99, LiLo99, Lin00, AlLo00,
DeHo00, MeHo00}, integral equations \cite{Loz83, OuBh83, KjMa84,
GoLo85, KjMa_1, KjMa_2, KjMa_3, LoHe86, LoDi90, OuBh91, DaBr9597,
GrKj98, DeAn01}, density functional theories \cite{Gro90, PeNo90,
TaSc92ab, PeJo93, DiTa99, BaDe00}, field theoretical calculations
\cite{PoZe88, CoDu92, NeOr00, MoNe00, Net01}, or other approaches
\cite{HaLi97, NyHa99, RoBl96, SoCr99, Shk99a, Shk99b, PeSh99, ArSt99,
ArLe00, DiCa01, LaLe00, LaPi01}.  An excellent recent summary can be
found in Refs.~\cite{Bel00, JoWe01}.  Furthermore, experiments have
shown that DNA (a stiff, highly negatively charged polyelectrolyte)
can be condensed by multivalent counterions \cite{WiBl79, BlWi80,
WiBa80, RaPa92, PoRa94, Blo96}.  This correlation-induced attraction
is for instance believed to be important for the compaction of DNA
inside viral capsids \cite{LamLe00, GeBr00}.

Even though many of the abovementioned theories are quantitatively
very successful, they are often also mathematically fairly
involved---like any systematic attempt to improve upon mean-field
theory.  It is thus desirable to have a theoretical description which
-- based on some knowledge about the nature of the dominant
correlations -- gives a more direct insight into the physics.  For
example, the Wigner-crystal theories, starting with
Refs.~\cite{RoBl96, Shk99a, Shk99b, PeSh99}, follow these lines and
estimate the excess correlational free energy by looking at the
ordered ground state of a two-dimensional layer of adsorbed ions.  The
success of these semi-empirical approaches clearly depends on how well
one understands the existing correlations.  However, even though the
\emph{effects} of correlations on, say, effective potentials or the
phase behavior have been well studied in the past, their \emph{nature}
has attracted much less attention \cite{KjMa_3, RoBl96, AlDa98,
AlLo00}.

Recently several theoretical models have been published which
discretize the distribution of condensed ions on DNA (by assuming
occupiable lattice-sites) and treat the resulting partition function
analytically or numerically \cite{SoCr99, ArSt99, ArLe00, DiCa01}.
These works provided a further important step in understanding the
origin of attraction and the nature of the correlations involved, but
it is not always obvious how the employed discretization influences
their strength.

In the present paper we use Molecular Dynamics (MD) simulations in
order to study the nature of counterion correlations around charged
cylindrical polyelectrolytes (modeled as charged rods) in the absence
of additional salt and on the level of a dielectric continuum
approximation for the solvent.  We will investigate both the case of a
bulk system of parallel rods, in which the occurrence of a negative
pressure is a sure indicator of attractions, as well as a single pair
of rods, in which the total force between the macroions is the
appropriate observable.  Recall that we invariably are concerned with
\emph{effective} free energies of interaction after ``integrating
out'' ionic degrees of freedom, which generally renders these
interactions non pairwise additive.  Hence, our two sets of data
complement each other.  In both cases it provides additional insight
to look separately at contributions coming from electrostatic and
non-electrostatic origin.

Finally we would like to remind the reader that short-ranged
correlation induced attractions have to be carefully distinguished
from the much longer ranged attraction between like-charged macroions,
indications of which have been found experimentally in dilute
suspensions of highly charged colloids at low ionic strength (for
recent work see for instance Ref.~\cite{TaRa92, ItYo94, TaYa97,
YoYa99}).  Suggestions for a theoretical explanations of this
phenomenon rest on reconsiderations of DLVO theory \cite{SoIs84,
RoHa97, RoDi99, War00}, but the situation is much less
clear-cut---both experimentally \cite{PaWu94} as well as theoretically
\cite{Ove87, GrRo01, DeGr02}.  In the present article we will not be
concerned with these phenomena.


\section{The Simulation Method}\label{sec:simulation}

Simulation details for the bulk systems have been described in
Refs.~\cite{Des00,DeHo01,DeHo02}.  In brief: We perform MD simulations using
a Langevin thermostat to drive the system into the canonical state
\cite{GrKr86}.  We will need only two kinds of basic interactions.
First, a short range repulsion which we model by the repulsive part of
a Lennard-Jones (LJ) potential
\begin{equation}
  \hspace*{-0.5em}
  V_\romev(r) = 
  \left\{
  \begin{array}{c@{\quad:\quad}c}
    4\,\epsilon
      \left[\left(\frac{\sigma}{r}\right)^{12} -
            \left(\frac{\sigma}{r}\right)^6 +
            \frac{1}{4} \right] & r < 2^{1/6}\sigma \\
    0 & \text{otherwise}
  \end{array}
  \right.,
  \label{eq:WCA}
\end{equation}
where $\sigma$ is essentially the distance below which strong
repulsion sets in; we will use it as our unit of length.  The energy
scale is set by $\epsilon$, but for a purely \emph{repulsive}
LJ-potential its precise value will not matter and we set it equal to
the thermal energy $k_\romB T$.  Eqn.~(\ref{eq:WCA}) is sometimes also
referred to as the ``Weeks-Chandler-Andersen potential''
\cite{WeCh71}.  Second, the bare Coulomb potential
$V_\romC(r)$ between two charges $z_1 e$ and $z_2 e$ can be written as
\begin{equation}
  \beta e \, V_\romC(r)
  \; = \;
  z_1 z_2 \, \frac{\ell_\romB}{r},
\end{equation}
where $\beta\equiv1/k_\romB T$ and the Bjerrum length $\ell_\romB =
\beta e^2 / 4 \pi \varepsilon_0 \varepsilon_\romr$ measures the
coupling strength by specifying the distance at which two unit charges
have interaction energy $k_\romB T$.  For instance, using the relative
dielectric constant of water $\varepsilon_\romr \approx 80$ and
ambient temperature $T\approx 300\,\text{K}$ we have
$\ell_\romB\approx 7\,\text{\AA}$.  Under periodic boundary conditions
the total Coulomb energy is obtained by a sum over all pairs --
including the images --, for which we use efficient particle-mesh
routines \cite{HoEa88, DeHo98}.

Our system consists of immobile charged rods and mobile oppositely
charged counterions.  The rods are assembled from a string of
negatively charged spheres (``monomers'') of diameter $\sigma$ which
sit on a line at a separation of $b = 1.042\,\sigma$.  We place one
rod along the main diagonal of the central simulation box which under
periodic boundary conditions yields a hexagonal array of infinitely
long rods.  The counterions are also modeled as spheres of radius
$\sigma$.  If there are $N$ counterions of valence $v$, global charge
neutrality requires $\sqrt{3}L/b = N v$ and thus an average counterion
density of $n=N/L^3=\sqrt{3}/vbL^2$.  No simulations presented in this
paper contain additional salt.  Let us finally introduce the
dimensionless charge parameter $\xi = \ell_\romB/b$, which measures
the number of unit charges along one rod per Bjerrum length.  We
briefly remind the reader of the concept of Manning condensation,
stating that if $\xi v > 1$, a fraction $1-1/\xi v$ of counterions
will associate with the rod \cite{Man69}.  A precise meaning of
``association'' is provided by the Poisson-Boltzmann solution of the
cylindrical cell model \cite{FuKa51, AlBe51} and has been carefully
discussed in the literature \cite{KlAn81, LeZi84, BeDr84, DeHo00},

In a second set of simulations we place two rods parallel to the
$z$-axis of the cubic box.  Their separation is small compared to the
box length in order to decouple the pair of rods from their periodic
images.  In this case we used both three-dimensional periodicity as
well as only one-dimensional periodicity along the directions of the
rods.  The latter case was handled by a one-dimensional version of an
algorithm that has been termed ``MMM'' \cite{StSp01}.  This is an
alternative method for evaluating electrostatic forces in
three-dimensions based on a convergence factor approach of order
${\mathcal O}(N\log N)$, and which can be adapted straightforwardly to
two- and one-dimensional periodic systems.  We evaluate the forces
pairwise using two different formulas, one very efficient for particle
pairs with large distances in the non-periodic plane, the second
formula, which is slower, for particle pairs with small distance.  The
formulas are derived similarly to the formulas presented in
\cite{ArHo01a, ArHo01b}.  The formula for distant particle pairs uses
a Fourier transform, while the formula for near particles uses an
expansion in polygamma functions. In this complementary approach we
also used a continuous line charge on the rod axis instead of an
aligned assembly of point charges.  We found identical results when
using the three-dimensional and the MMM-based approach and will thus
in the following not distinguish between results originating from
either one.

We fixed the surface-to-surface distance of the two
rods to $2\,\sigma$, which is a typical separation at which
attractions can be expected at sufficiently high Bjerrum length.  For
these systems we considered three different rod diameters $r_0$,
namely $r_0/\sigma = 0.5$, $2.0$ and $6.271$, which were implemented
by the repulsive part of a cylindrical Lennard-Jones potential, which
was shifted towards larger radii by replacing $r$ in
Eqn.~(\ref{eq:WCA}) by $r-(r_0-\sigma/2)$.  The system 3 with the
largest radius has a ratio of rod radius to charge spacing of 6.0,
which is similar to that of DNA, where it is $10\text{\AA} /
1.7\text{\AA} \approx 5.9$, hence we refer to it as the DNA-like
system.  The system parameters can be found in
Table~\ref{tab:systems}.

Let us finally remark that the assumption of a homogeneously charged
rod or linearly aligned point charges is often only a first
approximation, since many important stiff polyelectrolytes (\eg, DNA
or actin filaments) feature a \emph{helical} charge distribution.
Implications of this additional structure on many aspects of
helix-helix interactions are discussed in detail in a series of
theoretical papers by Kornyshev and Leikin \cite{KoLe, KoLe02} (see
also the simulations in Ref.~\cite{GuNi86,AlLo00}).  Since in the present
work we are concerned with more generic questions about the origin of
correlations, we will neglect this complication.

\begin{table}
  \begin{ruledtabular}
  \begin{tabular}{ccccc} 
  system & $r_0/\sigma$ & $b/\sigma$ & $s/\sigma$ & $L/\sigma$ \\
  \hline
  \textbf{1} & 0.50 & 1.042  &  2.95 & 134.5  \\
  \textbf{2} & 2.00 & 0.1538 &  6.00 &  42.00 \\
  \textbf{3} & 6.27 & 1.042  & 14.54 & 134.5  \\
  \end{tabular}
  \end{ruledtabular}
  \caption{Geometry of the simulated two-rod-systems. $r_0$ is the rod
  radius, $b$ the separation of unit charges along the rod, $s$ the
  axial separation of the rods and $L$ the box length.  The
  surface-to-surface distance of the rods is $1.95\sigma$ in case 1
  and $2\,\sigma$ in the other two cases.  The ions are always
  trivalent.}\label{tab:systems}
\end{table}


\section{Bulk systems}


\subsection{Osmotic coefficient}

The osmotic coefficient $\oscoeff$ is defined as the ratio between the actual
pressure and a fictitious pressure that would act if all interactions were
switched off (``ideal gas'').  For instance, the osmotic pressure of a dilute
solution of charged rods is overwhelmingly dominated by the counterions, but a
certain fraction of them may be sufficiently strongly localized by the
macroions such that they do not contribute their full share (one says they are
``osmotically inactive'').  In the regime of counterion condensation, $\xi v >
1$, the osmotic coefficient should approach $1/2\xi v < 1$ in the infinite
dilution limit, while for finite concentrations $\oscoeff$ is larger
\cite{Man69, FuKa51, AlBe51, KlAn81, LeZi84}.

\begin{figure}
  \includegraphics[scale=0.8]{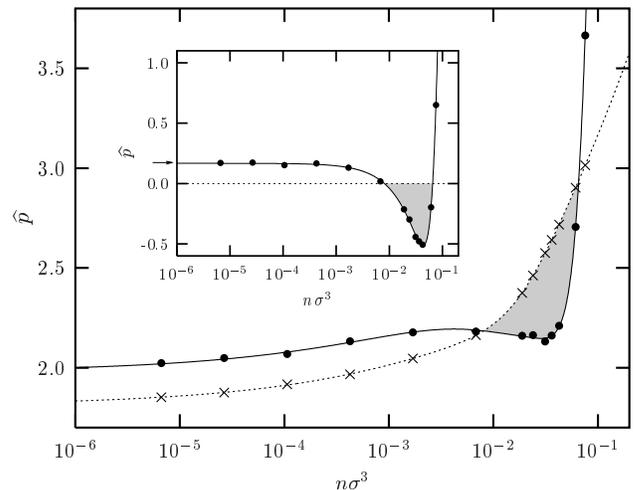}
  \caption{Contributions to the osmotic coefficient $\oscoeff$ as a
  function of counterion density $n$ for a bulk systems of parallel
  charged rods characterized by $r_0/\sigma=1$, $b/\sigma=1.042$,
  $\ell_\romB/\sigma=1$, and $v=3$.  The heavy dots and crosses
  represent the non-electrostatic contribution (kinetic and excluded
  volume) and the negative of the electrostatic contribution,
  respectively (the curves are guides to the eye).  The inset shows
  the total osmotic coefficient, and the limiting value of infinite
  dilution $1/2\xi v \approx 0.174$ is indicated by an
  arrow.}\label{fig:oscoeff}
\end{figure}

Figure~\ref{fig:oscoeff} shows our MD-results for the osmotic
coefficient of bulk systems which are characterized by $r_0/\sigma=1$,
$b/\sigma=1.042$, $\ell_\romB/\sigma=1$, and $v=3$.  The pressure is
identified with the component of the stress tensor perpendicular to
the rods \cite{DeHo01}, and its contributions coming from
electrostatic and non-electrostatic (\ie, entropic and excluded
volume) origin are plotted separately.  It can be seen that within a
density range from roughly $n = 8 \times 10^{-3}\sigma^{-3}$ to $6.5
\times 10^{-2}\sigma^{-3}$, corresponding to rod separations between
$6.8\,\sigma$ and $2.4\,\sigma$, the osmotic pressure is negative. If
the rods were not forced to remain at fixed separations, they would
phase separate, producing a condensate and a dilute phase.

Surprisingly, the electrostatic contribution (which is always negative
and always favors contraction) shows no particular features in the
regime where $\oscoeff < 0$.  It rather appears that the negative
pressure originates from a drop in the repulsive excluded volume
contribution which is large enough such that the electrostatic
contribution can win against the short range repulsion.  This finding
is consistent with earlier simulation results \cite{NiGu91} which used
finitely replicated cells.  A possible explanation of this phenomenon
rests on the following tempting picture: The excluded volume
contribution to the pressure stems from the force that ions exert on
the oppositely charged rods.  If the density increases, the rods come
closer to each other and start to pull the ions away from their
neighbors, thereby reducing this force.  However, at too high
concentrations rods will again repel by pushing onto each other via
the ions in between.  We will come back to this effect in
Sec.~\ref{ssec:azimutal}.

An alternative explanation suggests that the condensed ions between
two neighboring rods form a mutually interlocked pattern, which
results in a lower Coulomb energy (compared to a homogeneous charge
distribution) and which thus leads to attractions \cite{RoBl96,
GrMa97, ArLe00, LeAr99}.  Since each rod has six neighbors, this
defines six ``planes'' in which one would have to look for a
two-dimensional interlocking pattern.  However, we were unable to find
a corresponding structure in our simulated data that goes beyond a
weakly developed first correlation hole, which on its own is not a
sure sign for attractions as we will show in Sec.~\ref{ssec:interlock}
for the case of two rods.  This may partly be related to the small
degree of arbitrariness in the definition of the planes, but there is
also a physical explanation to consider.  The systems are
comparatively dense in the regime where attractions occur.  It is
therefore likely that close ions are correlated irrespective of the
plane they ``belong'' to---in other words, that those artificial
planes are irrelevant for understanding the physics.  It rather seems
more reasonable to study the full three-dimensional correlations of
the ions, as we will do in the following section.


\subsection{Pair correlation function}\label{ssec:3dpair}

If the radius of the rods and their mutual distance is smaller than
the Bjerrum length, the electrostatic interaction of ions reaches
beyond the nearest rod and the complete system of ions may form a
three-dimensional correlated structure.  More specifically, Shklovskii
suggests \cite{Shk99a} that in the case $r_0/b \ll v$, or equivalently
$r_0/\ell_\romB \ll v/\xi$, ions could form such a structure on the
background of the rods, which for very high coupling can be described
as a three-dimensional Wigner crystal (or at least as a strongly
correlated liquid).  We tested this assumption by computing the usual
pair correlation function $g(r)$ for the systems above which showed
negative pressure.  Before we discuss the results, a few general
remarks are appropriate.

If the ions form a three-dimensional crystal structure, there will be
a typical minimum distance between them which should scale like one
over the third root of the counterion density \footnote{Note that this
is a consequence of the long range interactions. In a system of hard
spheres the correlations become stronger at increasing density, but
there the first maximum of $g(r)$ is always at contact, \ie, its
position is independent of density.}.  However, there is a second
important length scale, which is the separation of the rods and which
is easily seen to scale as one over the square root of the counterion
density.  Hence, whatever the details of the actual ionic structure
are, if one measures a characteristic length varying like $n^{-1/3}$,
it can be viewed as a bulk correlated liquid property, while a length
varying like $n^{-1/2}$ is rod-imposed.

\begin{figure}
  \includegraphics[scale=0.8]{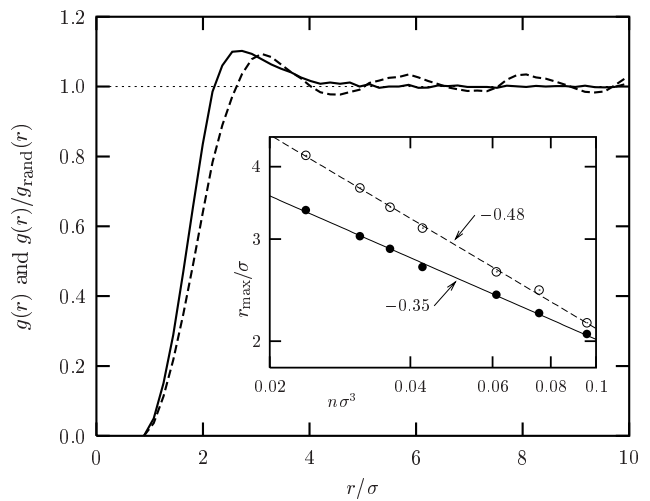}
  \caption{Pair correlation function $g(r)$ for the bulk system from
  Fig.~\ref{fig:oscoeff} at a counterion density $n = 4.25 \times
  10^{-2} \sigma^{-3}$ (dashed line). The solid line is the ratio
  between this $g(r)$ and $g_\rand(r)$ (see text). The inset shows a
  double logarithmic plot of the position of the first maximum of
  $g(r)$ (open circles) and $g(r)/g_\rand(r)$ (full circles) as a
  function of density; the slopes are indicated.}\label{fig:g_r}
\end{figure}

In Fig.~\ref{fig:g_r} we present results for the system studied above
in the density range $0.02\,\sigma^{-3} \ldots 0.1\,\sigma^{-3}$,
which corresponds to the high-density range in which the osmotic
pressure was found to be negative.  The dashed curve shows the pair
correlation $g(r)$ at the density $n = 4.25\times 10^{-2}
\sigma^{-3}$, which is at the minimum of the osmotic coefficient in
Fig.~\ref{fig:oscoeff}.  A pronounced oscillatory structure is clearly
visible.  The inset shows the position of the first maximum as a
function of density in a double logarithmic plot (open circles).  In
this range the data suggest a power law with exponent $-0.48$, which
is close to the value expected for a rod-imposed structure.  In fact,
the oscillations in $g(r)$ simply reflect the periodicity of the array
of rods.

In order to extract actual \emph{inter-ionic} correlations from the
simulations, the imposed periodic inhomogeneity of the ion density has
to be removed.  One way of doing this is as follows: The pair
correlation function is the probability of finding an ion at a
distance $r$ from another ion relative to the probability of finding
an ion at the same distance in a random (\ie, noninteracting, ``ideal
gas'') system at the same average density.  In the present case it
would be sensible to normalize not by a random homogeneous system but
by a random \emph{inhomogeneous} system, \ie, a system in which there
are no inter-ionic correlations but the spatially varying ion density
is preserved.  This can be easily accomplished in the following way:
In each configuration move every ion a random distance (between $0$
and $\sqrt{3}L$) along the direction of the rods.  This does not
change the inhomogeneous one-particle distribution, but completely
destroys all two-particle correlations.  Then compute the pair
correlation function of this randomized system, $g_\rand(r)$.  The
ratio between the usual $g(r)$ and the randomized $g_\rand(r)$ now
contains all information about inter-ionic correlations, but the
imposed density inhomogeneity is divided out.  This ratio is also
plotted in Fig.~\ref{fig:g_r} for the system at the density $n =
4.25\times 10^{-2} \sigma^{-3}$ (solid line).  Signs of correlations
can again be seen, but the long range oscillatory part has been
removed.  The inset also shows the position of the first maximum of
$g(r)/g_\rand(r)$ as a function of density (closed circles).  The
solid line has a slope of $-0.35$, which is more close to the exponent
expected for a correlated three-dimensional ionic structure. 

We see that the nature of correlations in this system is a subtle
interplay between rod-imposed and inter-ionic contributions, and the
latter are only identifiable once the former are divided out.  The
idea that a three-dimensional pattern of correlated ions forms on the
structureless background of the rods is clearly too simple for the
system under study.  Furthermore, both correlations are comparatively
weak, in the sense that the first maximum in $g(r)$ is fairly low, and
the long range structure visible in the bare $g(r)$ is identical in
$g_\rand(r)$ and thus is imposed by the external periodicity of the
rods.  The observed correlations are definitely much weaker than
expected for a Wigner crystal or a strongly correlated liquid.  If the
ideas advocated in Ref.~\cite{Shk99a, Shk99b, PeSh99} about the source
of attraction are correct, the present analysis implies that a
\emph{very} low degree of correlations is already sufficient or, in
other words, the Wigner crystal picture holds way beyond the ground
state.  This of course provokes the questions ``Why?'' and ``How far
beyond?''.

The above analysis does not directly explain our earlier finding that
the occurrence of a negative pressure is ultimately related to a
sudden drop in the repulsive excluded volume forces.  One may
speculate that the electrostatic part of the pressure responds less
sensitively than the short range LJ-part to a comparatively local
ordering as observed in Fig.~\ref{fig:g_r}, but additional studies
would be needed to support this.


\section{One pair of rods}

The Wigner crystal picture is ultimately based on energetic arguments,
it does not provide a direct ``mechanical'' explanation for why
correlations actually produce an attractive force.  To answer this
question, these correlations have to be studied in more detail.
However, the bulk system is not necessarily the easiest case: The
relevant observable is the pressure, which is less direct than a
force, and the more complex geometry complicates the definition of
suitable and reasonably intuitive correlation functions.  We therefore
resort now to the interaction between two rods as well as the involved
ionic correlations.  As we have mentioned in the introduction,
studying the pair interactions is not merely an alternative but also a
complementary approach to studying the bulk system, since the forces
in the latter cannot be decomposed into pair forces.

A pair of parallel rods has previously been investigated as a function
of rod separation, which gives the distance dependence of the force
\cite{GrMa97,ArSt99,AlLo00,DiCa01} and by integration the potential of
mean force.  In our work we supplement these results by studying the
properties of the system at \emph{fixed} distance as a function of
\emph{Bjerrum length}.  This alternative scan is promising since
unlike for bare charges the Bjerrum length does not simply enter as a
prefactor of the interaction.  Rather, its size will determine to what
degree ions can be found close to the rods and how strongly they
correlate with it as well as with each other, thereby influencing the
nature of the force itself.

We studied three systems which differ in their values for rod radius
and line charge density, as specified in Tab.~\ref{tab:systems}, but
we always kept the surface-to-surface distance between two rods around
$2\,\sigma$, \ie, twice the diameter of a counterions, since this is a
typical separation at which attractive forces have been found at
sufficiently large Bjerrum length.  The observables we measured were
the force between the rods (split again into electrostatic and
non-electrostatic origin) as well as various ionic correlation
functions along and around the rods.  The three studied systems will
provide examples for qualitatively different behavior, but cannot
predict detailed dependencies on key system parameters.  For instance,
the simulations in Ref.~\cite{GrMa97} suggest that increasing the rod
radius at fixed line charge density will ultimately remove repulsion,
while increasing the line charge density at fixed rod radius will
ultimately lead to attraction.  However, present studies in this direction
show a more complex behavior \cite{ArDe}.

It is well known that the importance of correlations is strongly
influenced by the valence of the counterions.  However, this variable
cannot be changed continuously.  In contrast, increasing the Bjerrum
length can be viewed as continuously increasing the strength of
correlations and thereby studying in detail how they develop.
Experimentally the Bjerrum length is a parameter whose change requires
some effort, but it can be achieved for instance by a careful control
of the solvent dielectric constant, as has been recently shown in an
experimental study of the coil-globule transition of DNA
\cite{MeKh99}.


\subsection{Force as a function of Bjerrum length}

In a first step we study the force between the two rods by
(separately) adding up all electrostatic and excluded volume forces
that act on one rod and average over the simulation run (which
typically means over about 3000 configurations separated by
2000 integration steps).  Under periodic boundary conditions
the total force on one rod originates not just from the neighboring
rod, but also from all periodic images.  However, we assume these
contributions to be negligible in our case for the following reasons:
For high Bjerrum length all ions are condensed onto the surface of the
rods, implying that interactions of these essentially neutral objects
(\ie, rods plus counterions) over a distance of the box length is much
weaker than the direct interaction (compare the values for $s$ and $L$
in Tab.~\ref{tab:systems}).  It is not immediately clear that this
remains valid in the limit $\ell_\romB
\rightarrow 0$, but in this case our data are always asymptotic to the
analytical expression for the electrostatic force between two charged
lines (see below), which implies that the images can be neglected as
well.

One may ask why we used periodic boundary conditions in the first place, if we
want images to be irrelevant.  The answer is of course that the images
\emph{along} the direction of the rod are important, since they render the
rods infinitely long---or stated differently: remove end effects.
Periodic replication perpendicular to the axis of the rods is an
unwanted but acceptable (since controllable) side effect imposed by
the use of traditional Ewald schemes, in which it is not trivial to
switch off periodicity in a specific direction.  However, the latter
can be achieved quite easily by using a one-dimensional version of MMM
\cite{StSp01,ArHo01a,ArHo01b}, compare Sec.\ref{sec:simulation}, and
we will also present data obtained by this second method.

Motivated by our observations in bulk simulations we also look at the
components of the force originating from electrostatic and excluded
volume interactions.  Let us start with a few general remarks
concerning what to expect.  For $\ell_\romB = 0$ there is no
electrostatic interaction between the rods, and the only possible
source of interaction is a depletion attraction \cite{AsOo54}
generated by the excluded volume of the (at $\ell_\romB = 0$
effectively uncharged!)  ions \footnote{The depletion force depends on
the density of ions, which should be zero if all we have is a pair of
rods.  In fact, our cells have of course a finite size and so the
ionic density is finite.  However, the depletion force it gives rise
to is essentially negligible.  For instance, for system 1 we find
$\beta F(\ell_\romB = 0)/L \approx -3\times 10^{-4}\sigma^{-2}$.}.
When slightly increasing the Bjerrum length, the two rods will feel an
unscreened Coulomb repulsion, since for $\xi v < 1/2$ ions will remain
unbound
\cite{Man69}.  The force per unit length is then given by
\begin{equation}
  \beta F \; \stackrel{\ell_\romB\rightarrow 0}{=} \; \frac{2
  \ell_\romB}{b^2 s}, \label{eq:F_ellB0}
\end{equation}
\ie, proportional to Bjerrum length. In other words, in this regime
the Bjerrum length still acts as a prefactor determining the strength
of the interaction.  Upon further increase of $\ell_\romB$ ion
condensation and subsequent correlations set in, resulting in
nontrivial $F$-$\ell_\romB$-curves (see below).  However, in the limit
$\ell_\romB \rightarrow \infty$ everything becomes simple again: The
counterions will ultimately assume a ``ground state'' configuration
which will no longer change upon further increase of $\ell_\romB$
(only fluctuations around the ground state will become weaker).
Hence, the electrostatic force (and as an indirect consequence also
the excluded volume force) will again be proportional to $\ell_\romB$.
However, the constant of proportionality cannot be predicted from
these considerations---not even its \emph{sign}.  In the following we
will plot the force per unit length and use the convention that a
\emph{positive} sign denotes \emph{repulsion}, while a \emph{negative}
sign denotes \emph{attraction}.

\begin{figure}
  \includegraphics[scale=0.8]{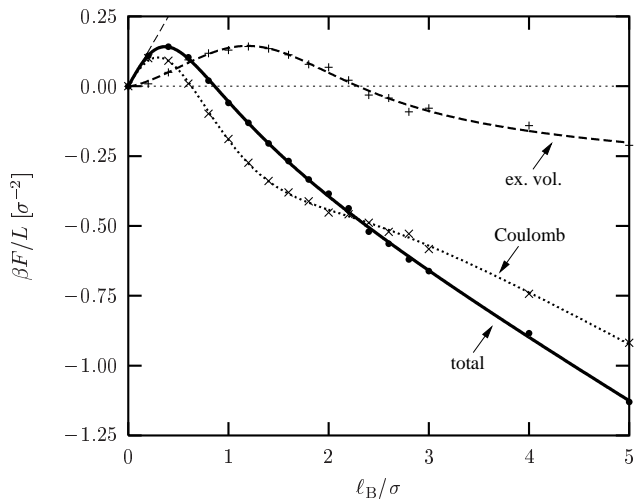}
  \caption{Force between two rods of system 1 from
  Tab.~\ref{tab:systems} as a function of Bjerrum length $\ell_\romB$.
  Total force, excluded volume contribution and electrostatic
  contribution are represented by ``\boldmath$\cdot$\unboldmath'' on a
  solid line, ``$+$'' on a dashed line, and ``$\times$'' on a dotted
  line, respectively.  The fine dashed line indicates the limiting
  behavior from Eqn.~(\ref{eq:F_ellB0}).  Positive forces denote
  repulsion, negative attraction.  Measured values are indicated by the
  symbols, the lines should merely guide the eye.}\label{fig:F_Bj_1}
\end{figure}

Fig.~\ref{fig:F_Bj_1} shows the result for system 1 (for notation see
Tab.~\ref{tab:systems}), which consists of rods having the same
diameter as the ions and the same separation as in the system from
Fig.~\ref{fig:oscoeff} showing the most negative pressure.  For small
Bjerrum length the force is given by Eqn.~(\ref{eq:F_ellB0}).
However, if $\xi v > 1/2$ (\ie, $\ell_\romB > 0.17\,\sigma$ the
present case) Manning condensation will set in on the two-rod system.
Since ions will condense preferentially \emph{between} the rods, they
reduce the electrostatic repulsion, but at the same time produce an
outward pressure due to their excluded volume.

Provided the excluded volume contribution to the interaction is not
yet substantial, a description of the system using Poisson-Boltzmann
(PB) theory can work up to this Bjerrum length.  Indeed, the PB
equation can be solved exactly in this geometry \cite{ImOn59,OhIm60},
but the analytical solution (in terms of hypergeometric functions) is
extremely complicated.  One particularly direct result however (exact
for infinitely thin rods) is that if the two-rod-system is below the
Manning threshold, pure Coulomb repulsion (\ie, unmodified by the
presence of counterions) is found asymptotically at large separation,
while if each rod is at the Manning threshold, the Coulomb repulsion
at large distances is reduced by a factor of 2.  We mention aside that
for the case of two rods (of arbitrary radius) and added salt the PB
equation has been solved numerically \cite{Har98}, and analytical
studies in the presence of salt exist within \emph{linearized} PB
theory \cite{BrMc73}, even for tilted rods \cite{BrPa74}.

Upon further increase in Bjerrum length the electrostatic force is
seen to change sign at $\ell_\romB / \sigma \approx 0.6$, and the
total force becomes attractive beyond $\ell_\romB / \sigma \approx
0.9$.  While it is easy to imagine counterion distributions between
the two rods that would lead to electrostatic attraction, those
distributions are counteracted by both entropy and excluded volume
interactions among the ions, since a high density between the rods is
required.  A mean-field treatment on the level of PB theory
\cite{ImOn59, OhIm60, Har98} cannot resolve the issue because rigorous
proves exist that the above situation must give repulsion \cite{Neu99,
SaCh99, Tri00}---contrary to the actual observation, which is in
agreement with earlier simulational works \cite{LyNo97, GrMa97}.  It
is clear that (on the level of the restricted primitive model of
electrolytes) the attraction must hence be related to the existence of
correlations between the ions.  We will discuss a few of them in the
following section.

For a Bjerrum length larger than $2.5\,\sigma$ the excluded volume
part of the force becomes attractive as well.  This is surprising,
since it implies that there are still sufficiently many ions between
the rods to induce electrostatic attraction, but they organize their
positions such that they put less pressure on the rods than the ions
located on the outward surfaces.

\begin{figure}
  \includegraphics[scale=0.8]{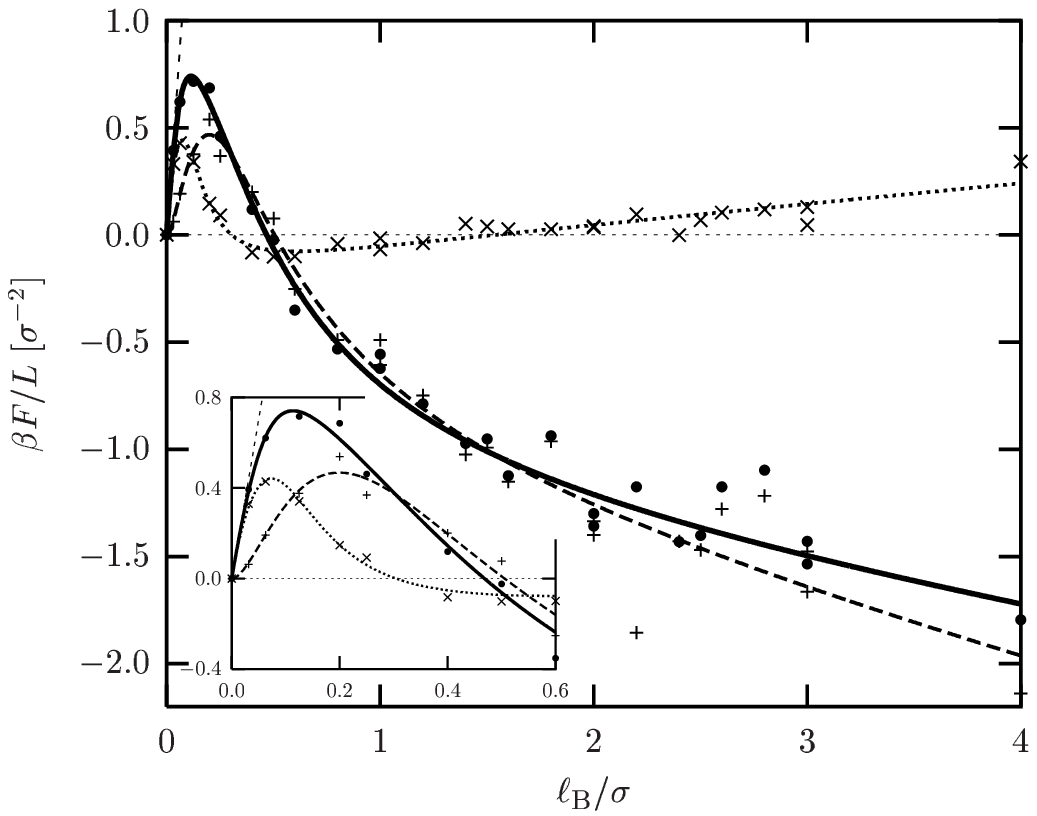}
  \caption{Force between two rods of system 2 as a function of Bjerrum
  length.  The inset magnifies the initial region
  $\ell_\romB\in[0;0.6]$.  The line styles are the same as in
  Fig.~\ref{fig:F_Bj_1}.}\label{fig:F_Bj_2}
\end{figure}

In Fig.~\ref{fig:F_Bj_2} we show the same Bjerrum-scan for system 2,
which differs from the previous one in the following ways: The linear
charge density is $6.78$ times larger, the rod radius is four times
larger, and the rods are kept at a separation of $6\,\sigma$, which
ensures that the surface-to-surface separation is again $2\,\sigma$.
The generic behavior at small Bjerrum length is the same, but in this
system a very pronounced difference occurs at larger values: Beyond
$\ell_\romB / \sigma \approx 1.5$ the electrostatic contribution to
the force becomes \emph{repulsive} and the total force is attractive
only because of the excluded volume term.  Note that together with the
results discussed above this implies that a net attraction can occur
both because electrostatic attraction overcomes excluded volume
repulsion and because excluded volume interactions overcome an
electrostatic repulsion.  In the present case electrostatic repulsion
is only weakly developed and numerical errors are large compared to
the other two systems, but we have observed the same phenomenon in
several other systems as well.  Its characteristics will be discussed
more thoroughly somewhere else \cite{ArDe}.  Even though this effect
may appear counterintuitive, we want to remind the reader that
minimizing the total energy of the ionic system at given rod
separation does not imply that the electrostatic \emph{part} of it
gives rise to rod-rod-attractions.

\begin{figure}
  \includegraphics[scale=0.8]{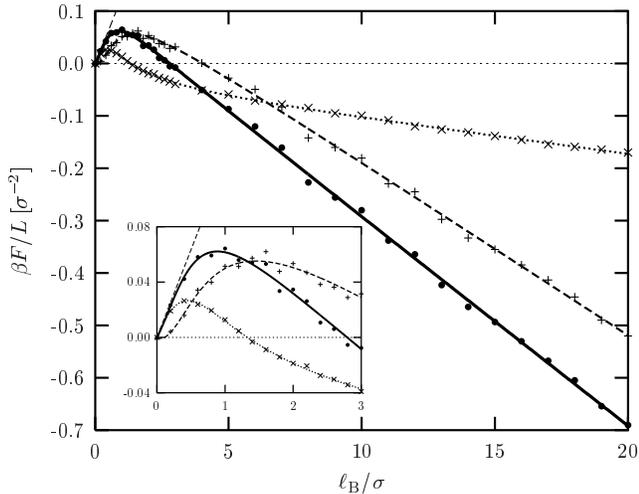}
  \caption{Force between two rods of system 3 as a function of Bjerrum
  length.  The inset magnifies the initial region
  $\ell_\romB\in[0;3]$.  The line styles are the same as in
  Fig.~\ref{fig:F_Bj_1}.}\label{fig:F_Bj_3}
\end{figure}

Fig.~\ref{fig:F_Bj_3} presents results for system 3, which compared to system
1 has an approximately $12.5$ times larger rod radius.  The
sur\-face-to-sur\-face separation is again maintained at $2\,\sigma$.  The
ratio $r_0/b \approx 6$ is close to that for DNA. Fixing the length scale via
$b=b_{\text{DNA}}=1.7\,$\AA\ implies a (somewhat small) ion diameter of
$\sigma=1.63\,$\AA\ \footnote{However, individual ions still have sufficiently
  large distances from each other, implying that the excluded volume
  interaction \emph{between ions} is not yet significant. Hence, the actual
  ionic size does not matter too much.}.  At the Bjerrum length $\ell_\romB =
7.14 \, \text{\AA} = 4.38 \, \sigma$ (as appropriate for water at room
temperature) the total force is attractive, and indeed it has been
experimentally established that trivalent ions lead to attractive interactions
between DNA strands \cite{WiBl79, BlWi80, WiBa80, RaPa92, PoRa94, Blo96}.
Note also that just like in system 1 both contributions to the force are
attractive at sufficiently high Bjerrum length; however, in system 3 the
excluded volume part is \emph{stronger} than the electrostatic part for values
of $\ell_\romB / \sigma$ larger than $\approx 6$.

\begin{figure}
  \includegraphics[scale=0.8]{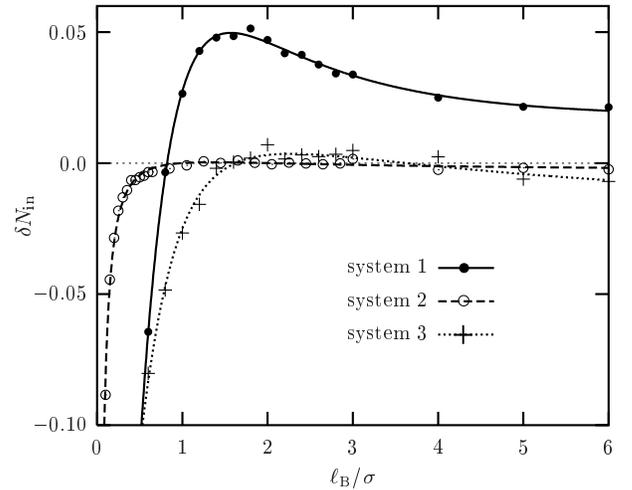}
  \caption{Relative imbalance $\delta N_{\text{in}} = (N_{\text{in}} -
  N_{\text{out}})/(N_{\text{in}}+N_{\text{out}})$ of ions between the
  rods and outside (see text) as a function of Bjerrum length for the
  three systems studied.  Note that a positive $\delta N_{\text{in}}$
  means that more ions are between the rods.}\label{fig:in_out}
\end{figure}

As we have mentioned above, the attraction between the rods due to
electrostatic forces arises because ions condense preferentially
\emph{between} the rods.  A straightforward measure for this imbalance is
provided by the following observable: Imagine two parallel planes, each
containing the axis of one of the rods, whose distance equals the axial
distance between the rods.  These planes divide space into a region
``between'' the rods and two disjunct regions ``outside''.  Denote by
$N_{\text{in}}$ and $N_{\text{out}}$ the number of counterions in the region
``between'' and ``outside'', respectively, and define the relative ionic
excess between the rods as $\delta N_{\text{in}} = (N_{\text{in}} -
N_{\text{out}})/(N_{\text{in}}+N_{\text{out}})$.  This observable is plotted
in Fig.~\ref{fig:in_out} as a function of Bjerrum length for systems 1, 2 and
3.  For small Bjerrum length $\delta N_{\text{in}}$ is negative, since ions
are not condensed and the outside region is larger.  As ion condensation sets
in at increasing $\ell_\romB$, $\delta N_{\text{in}}$ also increases until it
reaches 0, the point at which the inside and outside numbers are balanced.
For system 1, $\delta N_{\text{in}}$ becomes strongly positive afterwards,
showing that there are more ions between the rods than outside, but it
decreases again beyond $\ell_\romB/\sigma\approx 1.5$, implying that this
imbalance softens out.  For system 2, $\delta N_{\text{in}}$ does not rise
significantly above 0, but rather approaches this balance line at about
$\ell_\romB/\sigma \approx 1$.  For system 3, $\delta N_{\text{in}}$ weakly
rises above 0 at $\ell_\romB/\sigma \approx 1.6$, but it again drops below at
$\ell_\romB/\sigma \approx 3.6$.

In all three systems the point at which the electrostatic force
becomes attractive coincides roughly with the point at which a
strongly negative imbalance $\delta N_{\text{in}}$ vanishes.  However,
one has to be careful in interpreting this observable: For system 1
the excluded volume repulsion between the rods vanishes beyond
$\ell_\romB/\sigma\approx 2.5$, but $\delta N_{\text{in}}$ is positive
there, \ie, there are still more ions between the rods than outside.
The attraction between the rods in system 2 is due to excluded volume
forces, but the imbalance $\delta N_{\text{in}}$ is essentially 0.
Hence, attraction occurs not due to a simple density difference.  In
the following section we will therefore also study the azimuthal
distribution of ions in some more detail as well as the question how
close the ions actually come to the rods.

Let us close our discussion of forces with the following observation.
Compared to system 1 the density of surface charges $\varsigma$ is
$12.5$ times lower in system 3, but the total force becomes attractive
at $\ell_\romB / \sigma \approx 2.9$, which is only $3.2$ times larger
than for system 1.  Rouzina and Bloomfield~\cite{RoBl96} use the
two-dimensional plasma parameter $\Gamma_2 = \ell_\romB v^{3/2}
\varsigma^{1/2}$ as a measure for the strength of correlations and
estimate that the onset of attraction should be expected at $\Gamma_2
\approx 2$. This implies that the corresponding Bjerrum length scales
inversely proportional to the square root of the surface charge
density. For our simulations $\Gamma_2 \approx 2.58$, $1.83$ and
$2.35$ for systems 1, 2 and 3 at the onset of attraction, which agrees
remarkably well with their estimate.


\subsection{Azimuthal correlations}\label{ssec:azimutal}

\begin{figure}
  \includegraphics{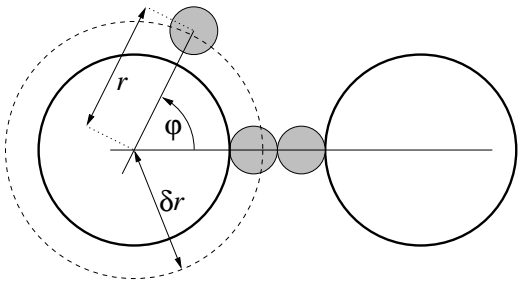}
  \caption{Geometry for the definition of the azimuthal correlation
  function $g(\varphi|\delta r)$. It is proportional to the
  probability density of finding a counterion at the angle $\varphi$
  relative to the other rod within a condensation distance of at most
  $\delta r$. The normalization is such that $g=1$ corresponds to the
  average density, or stated differently, the average of $g$ is
  $1$.}\label{fig:g_phi_def}
\end{figure}

The previous sections have revisited the importance of ionic
correlations for the nature of the force between two charged rods, and
in the following we will present measurements of a few of them.  We
start with an ion-rod correlation that answers the question how the
ions distribute around a rod relative to the position of the other
rod.  Fig.~\ref{fig:g_phi_def} illustrates our definition of the
correlation function.  Denote the position of a counterion relative to
rod 1 in polar coordinates $r$ and $\varphi$, such that $\varphi=0$
corresponds to the direction towards rod 2. We define
$g(\varphi|\delta r)$ to be $2\pi$ times the probability density of
finding an ion at the angle $\varphi$, given that its separation from
rod 1 is at most $\delta r$.  Note that this implies that the mean
value of $g(\varphi|\delta r)$ (when averaged over $\varphi$) is 1.
Obviously, $g(\varphi|\delta r)$ is periodic in $\varphi$ with period
$2\pi$.  At first we will choose $\delta r$ as large as possible, \ie,
half the rod separation $s$.  We want to mention that for $\delta r
\rightarrow r_0$ this function is essentially the rod-analog of
the density resolved by the polar angle for the spherical case studied
in Ref.~\cite{AlDa98}.

\begin{figure}
  \includegraphics[scale=0.55]{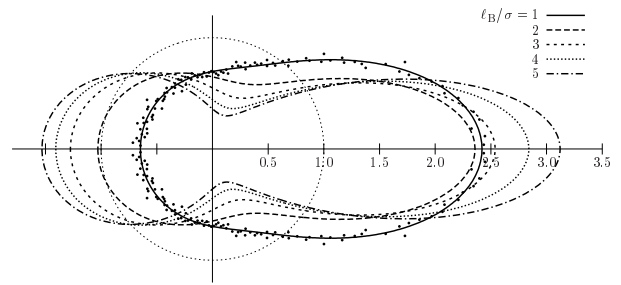}
  \caption{Polar plot of azimuthal correlation functions
  $g(\varphi|s/2)$ for system 1.  The particular Bjerrum lengths are
  indicated. The neighboring rod is assumed to be located to the right
  side. The normalization is such that the average value of $g$ is
  equal to 1 (indicated by the dotted circle). Measured values (dots)
  are only indicated for the case $\ell_\romB =
  1\,\sigma$.}\label{fig:azimut_generic}
\end{figure}

Figure~\ref{fig:azimut_generic} shows a polar plot of $g(\varphi|s/2)$
for system 1. The Bjerrum length is varied between $1\,\sigma$ and
$5\,\sigma$. Compare also the corresponding force-plot from
Fig.~\ref{fig:F_Bj_1}. Several things can be observed: ($i$) The
azimuthal correlation function is largest in the direction towards the
other rod. ($ii$) There exists a ``second tide'' on the side opposing
the other rod. ($iii$) The elongation of $g(\varphi|s/2)$ along the
line joining the two rods increases with increasing Bjerrum
length. ($iv$) The larger than average $g$ at $\varphi=0$ and maybe
$\varphi=\pi$ is balanced by a less than average $g$ in the transverse
direction. ($v$) While $g(0|s/2)$ is always larger than $1$,
$g(\pi|s/2)$ is smaller than $1$ for sufficiently small Bjerrum
length, but ultimately rises above $1$ as well.

As can be seen from Fig.~\ref{fig:F_Bj_1}, all five systems presented
in Fig.~\ref{fig:azimut_generic} feature a net attraction, but for the
systems with $\ell_\romB / \sigma = 1$ and $2$ the excluded volume
part is still repulsive.  It is unfortunately difficult to directly
relate this finding to the shape of the azimuthal correlation
function, since the position of the ions are of course relevant for
both electrostatic and excluded volume forces, which are opposite and
show a different distance dependence. We come back to this issue
below, when we discuss the dependence on $\delta r$.

\begin{figure}
  \includegraphics[scale=0.55]{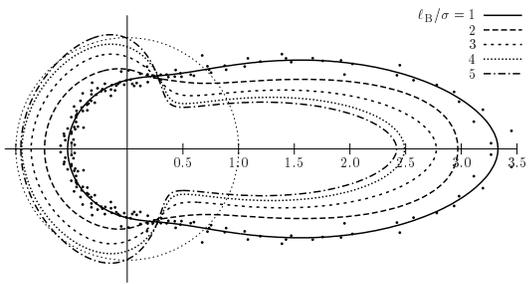}
  \caption{Same plot as in Fig.~\ref{fig:azimut_generic} for system 3.
  Measured values (dots) are only indicated for the case $\ell_\romB =
  1\,\sigma$.}\label{fig:azimut_large}
\end{figure}

Fig.~\ref{fig:azimut_large} shows exactly the same five correlation
functions for system 3.  This system differs from system 1 ``only'' by
a $12.5$ times larger rod radius, but the shape of the correlation
functions is quite different.  The second peak at $\varphi=\pi$ has
given way to two new peaks emerging somewhat beyond $\pi/2$, while the
ion density at $\varphi=\pi$ is below average. The fact that the peak
at $\varphi=0$ decreases upon increasing Bjerrum length can be traced
back to the normalization of $g$ and the fact that the total number of
ions within the shell $r\le s/2$ initially increases as $\ell_\romB$
increases.  At low Bjerrum length most of the condensed ions can be
found between the rods. Upon increasing $\ell_\romB$ the additionally
condensed ions will occupy the other parts of the rod, thereby
reducing the relative weight of the ones at $\varphi=0$. Once
essentially all ions have condensed within the shell $r\le s/2$, a
larger Bjerrum length will only enhance the correlations and thereby
make the peak at $\varphi=0$ larger again.  We note that in the spirit
of our mapping indicated above the azimuthal correlation function for
ions around DNA would approximately be given by the curves for
$\ell_\romB/\sigma = 4$ or 5.

\begin{figure}
  \includegraphics[scale=0.8]{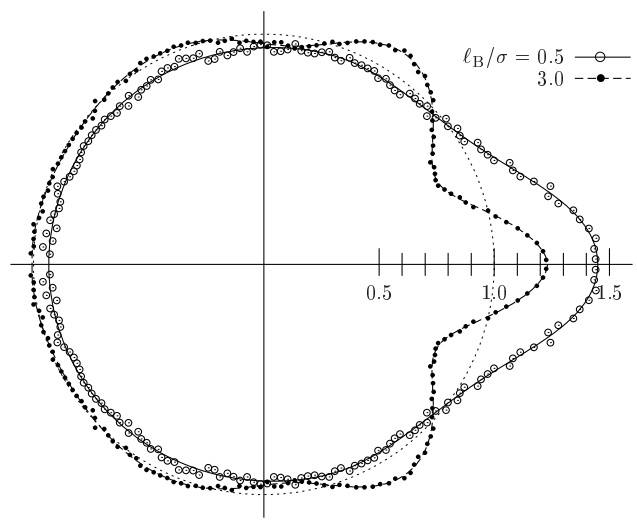}
  \caption{Same plot as in Fig.~\ref{fig:azimut_generic} for system 2.
  Measured values for $\ell_\romB/\sigma=0.5$ and
  $\ell_\romB/\sigma=3.0$ are indicated as open circles ($\circ$) and
  dots (\boldmath$\cdot$\unboldmath), respectively.  The lines should
  merely guide the eye.}\label{fig:azimut_DNA}
\end{figure}

In Fig.~\ref{fig:azimut_DNA} we show $g(\varphi|s/2)$ for two versions
of system 2. This system has a rod radius $4$ times larger than system
1 and a line charge density about 6.8 times larger.  For a Bjerrum
length of $\ell_\romB = 0.5\,\sigma$, for which the total force
between the rods approximately vanishes (see Fig.~\ref{fig:F_Bj_2}), a
pronounced peak in $g(\varphi|s/2)$ towards the neighboring rod as
well as a slightly smaller than average counterion density on the
opposite side is visible.  Just as for system 3, upon increasing the
Bjerrum length $g(\varphi|s/2)$ does not merely ``intensify'' its
features but develops a qualitatively new structure.  In addition to
the main correlation peak at $\varphi=0$, five new peaks show up,
roughly at multiples of $60^\circ$.  This occurrence of additional
peaks is a common phenomenon in usual liquids as correlations
increase. However, in the present case we have the additional
constraint that $g(\varphi|\delta r)$ is periodic in $\varphi$ and
obviously an even function, which restricts the locations of these
peaks. While for a \emph{flat} two-dimensional system the position of
peaks in the pair correlation function is solely determined by the
density of charges ($\varsigma^{-1/2}$ is the only available length
scale), in a \emph{curved} geometry two more effects play a role:
first, the radius of curvature appears as a new length scale, and
second, topological constraints require the function to close upon
itself as in the case above.  Particularly the latter observation
suggests that there are situations in which this matching works
``automatically'' whereas in other situations it leads to a
frustration. This may favor counterion conformations which relax these
constraints (\eg\ helices) and influence the strength of the
rod-rod-interaction.

\begin{figure}
  \includegraphics[scale=0.8]{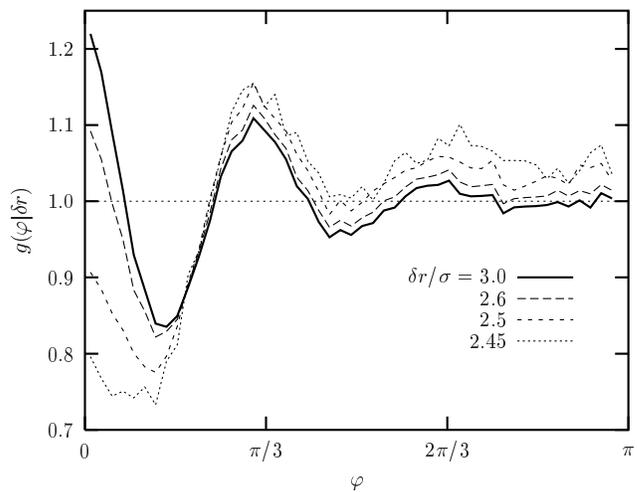}
  \caption{Azimuthal correlation functions $g(\varphi|\delta r)$ for
  the system 2 with $\ell_\romB=3\,\sigma$ (see also
  Fig.~\ref{fig:azimut_DNA}) for different values of $\delta r$. Note
  in particular that for decreasing $\delta r$ the peak at $\varphi=0$
  turns into a correlation hole.}\label{fig:azimut_dr}
\end{figure}

We conclude our discussion of the azimuthal correlation function by
discussing its dependence on $\delta r$. This variable determines
which counterions are taken into account for computing the density as
a function of $\varphi$. If $\delta r$ gets smaller, the focus is on
ions that are more closely in contact with the rod.
Fig.~\ref{fig:azimut_dr} shows a plot of $g(\varphi|\delta r)$ for
system 2 with $\ell_\romB=3\,\sigma$.  The bold solid line is the same
curve as shown in Fig.~\ref{fig:azimut_DNA} (\ie, $\delta r =
s/2=3\,\sigma$), only plotted in a Cartesian way.  The other three
curves correspond to successively smaller values of $\delta r$.  We
would like to draw attention to the fact that the peak at $\varphi=0$
gives way to a correlation hole.  While $g(0|\delta r)$ decreases upon
reduction of $\delta r$, $g(\pi|\delta r)$ increases.  This is
important since the ions most closely in contact with the rod produce
the strongest excluded volume force.  Fig.~\ref{fig:azimut_dr} thus
shows that ions very close to the surface of the rods are more likely
to push them inwards rather than outwards.  Note that this particular
system is special in that the attractive forces have been found to
originate from the excluded volume term, see Fig.~\ref{fig:F_Bj_2}. A
similar $\delta r$-analysis for system 1 in the strongly attractive
regime at $\ell_\romB = 20 \, \sigma$ does not show this effect (data
not shown).  As $\delta r$ gets smaller than $1.0\,\sigma$, the peaks
at $\varphi=0$ and $\varphi=\pi$ start decreasing and rising,
respectively, but the former does not turn into a correlation hole.
The same finding applies to system 3 in the attractive regime at
$\ell_\romB = 5 \, \sigma$.  However, note also that in the latter two
systems electrostatics is the dominant reason for attraction.
Unfortunately, effects at small $\delta r$ are difficult to observe,
since the small number of ions at these close distances impedes a good
statistics.

We want to remark that a similar correlation hole has been previously
observed in the spherical case and has been termed an
electrostatic depletion effect \cite{AlDa98}.  Observe that in our
case this effect is only visible if one focuses on ions very close to
the surface of the macroion---the \emph{total} amount of ions between
the macroions is \emph{above} average, which contradicts the depletion
picture.


\subsection{Ion interlocking}\label{ssec:interlock}

It has previously been suggested that the ionic correlations in the
case of attracting rods take the form of an interlocked pattern in the
plane between the rods \cite{RoBl96, GrMa97, ArLe00, LeAr99}. In this
section we present first measurements of these kind of correlations
under off-lattice conditions.  It turns out that alternating
charge-asymmetries indeed exist, but they are remarkably weakly
pronounced and fairly short ranged.  We also investigate their
relevance for the attraction.

\begin{figure}
  \includegraphics[scale=0.8]{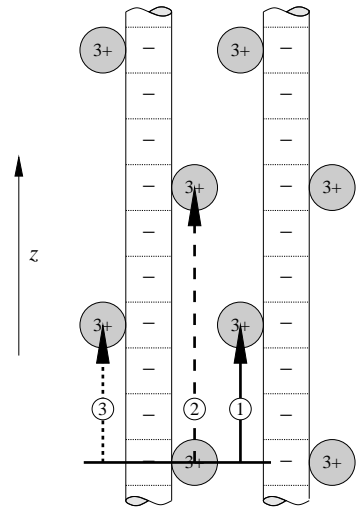}
  \caption{Scheme of a possible ``ground state'' of the two rod
  systems, showing ion interlocking. The relative position of three
  kinds of condensed counterions with respect to a condensed ion
  between the rods give rise to three different correlation functions,
  as discussed in the text.}\label{fig:interlock_ground}
\end{figure}

Fig.~\ref{fig:interlock_ground} envisages a tentative ``ground state''
for system 1 from Tab.~\ref{tab:systems}. We have previously deduced
from Fig.~\ref{fig:azimut_generic} that in the limit of high Bjerrum
length the ions are essentially either at $\varphi=0$ or $\varphi =
\pi$.  On each rod, every three monomers a trivalent counterion is
condensed.  The ions between the rods form an interlocked pattern that
also gets imprinted on the ions located on the opposite side of the
rods.  Such a pattern suggests the definition of the following three
pair correlation functions, as indicated in
Fig.~\ref{fig:interlock_ground}: Given an ion, that is condensed on
one rod and sits between the two rods, what is the probability of
having a second ion at a distance $z$ along the rods that is condensed
(1) on the opposing rod and also between the rods, (2) on the same rod
and also between the rods, and (3) on the same rod, but facing
outward?  With the usual normalization to $1$ at large distance, let
us call these three functions $g_1(z)$, $g_2(z)$ and $g_3(z)$,
respectively.  We only consider ions whose center has a distance of
less than $\sigma/2$ from the common plane of the rods.

\begin{figure}
  \includegraphics[scale=0.8]{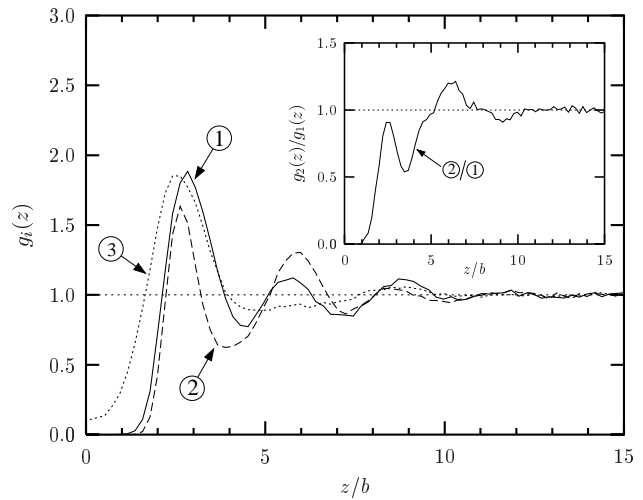}
  \caption{Interlocking correlation functions $g_1(z)$ (solid line),
  $g_2(z)$ (dashed line) and $g_3(z)$ (dotted line) as defined in
  Fig.~\ref{fig:interlock_ground} and the text for system 1 with
  Bjerrum length $\ell_\romB=5\,\sigma$.  The inset shows the ratio
  $g_2(z)/g_1(z)$.}\label{fig:interlock_generic}
\end{figure}

\begin{figure*}
  \includegraphics[scale=1.05]{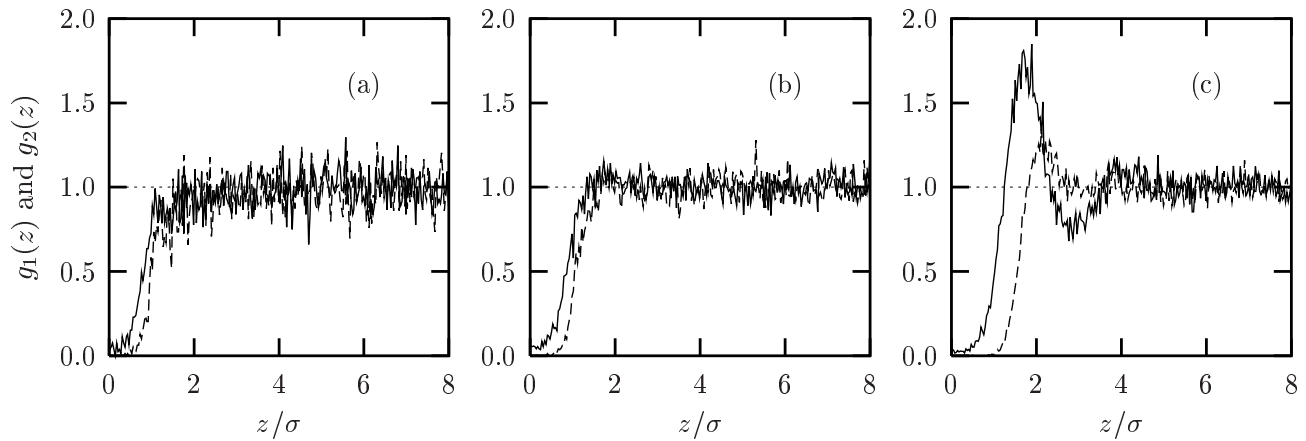}
  \caption{Interlocking correlation functions $g_1(z)$ (solid line)
  and $g_2(z)$ (dashed line) for system 2 with three values of the
  Bjerrum length: (a) $\ell_\romB = 0.125\,\sigma$, (b) $\ell_\romB =
  0.5\,\sigma$, and (c) $\ell_\romB =
  3\,\sigma$.}\label{fig:interlock_DNA}
\end{figure*}

Fig.~\ref{fig:interlock_generic} shows the result of a measurement of
these three correlation functions for system 1 with $\ell_\romB =
5\,\sigma$.  Pronounced correlations are clearly visible, but their
range is comparatively short.  Structural features in the correlation
function $g_3$ are much weaker developed.  The reason for this is that
the ionic density outside the rods is lower than between them, and the
structure arises from a weak ``transfer'' of the correlations of $g_1$
and especially $g_2$.

It may seem surprising that not just $g_1(z)$ but also $g_2(z)$ has
its first peak at $z \approx 3b$.  The reason for this is that just as
for the three-dimensional pair correlation function discussed in
Sec.~\ref{ssec:3dpair} a fair amount of long-ranged periodicity in the
$g_i(z)$ arises due to an external imprinting.  Here it is the fact
that charge neutrality essentially requires 1 trivalent ion per rod
every $3b$.  A visual inspection of typical ion configurations also
shows that the system is still far from a highly ordered ground state
as envisaged in Fig.~\ref{fig:interlock_ground}.  In particular, for
$\ell_\romB/\sigma=5$ the imbalance $\delta N_{\text{in}}$ presented
in Fig.~\ref{fig:in_out} has a value of about 0.022, which implies
that the average density between the rods is about 5\% higher than
outside.  These additional ions have to be put somewhere and clearly
destroy the simple groundstate from Fig.~\ref{fig:interlock_ground}.
Together these findings imply that $3b$ is also a typical distance
that has to be expected for $g_2$, but observe also that the peak in
$g_2(z)$ at $z\approx 3b$ is less pronounced than the one for
$g_1(z)$.  In fact, the terminology ``interlocking'' suggests that we
should be interested in the \emph{relative} charge asymmetries along
the rods.  A better measure would therefore be the \emph{ratio}
between the two inner correlation functions, as is shown in the inset
of Fig.~\ref{fig:interlock_generic}.  For $z \rightarrow 0$ the ratio
$g_2(z)/g_1(z)$ approaches 0, since the correlation hole of $g_2(z)$
at $z=0$ is of course more pronounced than that for $g_1(z)$.  The
small peak shortly before $z=3b$ indicates the high probability for
this distance as discussed above, but the fact that the peak value is
below 1 means that such a value occurs more likely for ions belonging
to different rods.  A somewhat broader peak can be seen around
$z\approx 6b$.  It has its value above 1, indicating that this
distance occurs preferentially for ions on the same rod.  Beyond
$z=10b$ no significant structure is visible.

Together these findings show that ion interlocking does indeed occur,
but that it is again superimposed by ``trivial'' periodicities along
the rods and that its extension along the rods is fairly weak.  For
increasing Bjerrum length the above features get more pronounced and
the oscillatory structure extends towards larger separations, while
for decreasing Bjerrum length the features diminish.  At $\ell_\romB =
1\,\sigma$, at which the total force is approximately $0$, the only
remaining feature of all three correlation functions is the
correlation hole at $z=0$ (data not shown), which survives at even
smaller values for $\ell_\romB$.

We would like to emphasize that this correlation hole at $z=0$ is not
sufficient to entail attractive forces. This is important for the
following reason: In the Wigner crystal approach the correlation
energy of an ion can be well approximated by its interaction energy
with the counter-charge within its correlation hole, which turns out
to give the dominant contribution to the energy up to temperatures far
 beyond the crystallization \cite{Shk99b}. This may give the
impression that the existence of this first correlation hole is also
sufficient for the appearance of attractive forces.
Fig.~\ref{fig:interlock_DNA} gives another illustration that this is
not the case.  The correlation functions $g_1$ and $g_2$ are plotted
for system 2 for three values of the Bjerrum length $\ell_\romB/\sigma
= 0.125$, $0.5$ and $3$.  As can be seen from Fig.~\ref{fig:F_Bj_2},
the first case corresponds to the strongest total repulsion, the
second to an approximately vanishing net force and the third case to
attraction. While the correlation functions in the first two cases are
very similar, in the third case pronounced interlocking is observed.

One word of caution should be addressed to the numerous theories which
apply the interlocking picture to DNA-like systems. Our simulations
show that the net attractive force in System 3 around $\ell_\romB
/\sigma \approx 4-5$ is only weak compared to the other two systems,
and thus an interlocking pattern at theses values of $\ell_\romB$ is
basically invisible.


\section{Summary, questions and outlook}

We have revisited the phenomenon of attractive interactions in systems of
charged rods in the presence of multivalent counterions.  Within bulk systems
this effect is seen as a negative osmotic pressure.  Separating between ($i$)
the electrostatic and ($ii$) the entropic plus excluded volume component
unveils the key role played by the short range repulsive forces.  We further
explored these components by measuring their contribution to the force between
two parallel charged rods as a function of Bjerrum length $\ell_\romB$.  An
important result is that an overall attraction can be due to $(i)$ both
electrostatic and excluded volume forces bringing the rods together, ($ii$) an
electrostatic attraction overcoming excluded volume repulsion, and ($iii$)
a short range excluded volume force pushing the rods together against an
electrostatic repulsion.  The excluded volume induced attraction can 
occur even if the average density between the rods is higher than outside,
hence it is not a simple depletion mechanism. This suggests that there exists
more than one mechanism for attraction between rods.  It is the subtle
interplay between several effects which ultimately determine the total force.
We have shown that this crucially depends on the parameters of the system,
like the line charge density or the rod radius.

We also presented measurements of various ionic correlation functions.  For
the bulk systems the scaling of the first peak of the three-dimensional pair
correlation function indicates the existence of a three-dimensional correlated
liquid, but only if one corrects for trivial effects due to the inhomogeneous
average ion density.  The assumption of a correlated liquid forming on the
background of the rods is thus too simple.  If the attractive interactions are
still to be explained on the basis of the weak and masked three-dimensional
correlations, this would imply that the Wigner crystal idea remains useful far
beyond the ground state.

For the two rod case we studied the distribution of ions around one
rod relative to the other.  We always found a correlation peak towards
the other rod, but the rest of the distribution depends sensitively on
Bjerrum length, line charge density and the rod radius.  On the far
side of the rod there exists a correlation hole at low Bjerrum length,
which may turn into a peak at increasing Bjerrum length.  Also,
secondary peaks can appear at a certain angle with respect to the
rod-rod direction.  The periodicity of the azimuthal correlation
function implies commensurability constraints, which influence the
structure.  The azimuthal correlation function was also shown to
depend on the condensation radius $\delta r$.  It has been
demonstrated for one system that the peak at $\varphi=0$ can turn into
a correlation hole if one focuses on the ions most close to the rod.
While this is important for the direction of the short range repulsive
forces, it is presently unclear under which circumstances this effect
occurs.

We finally confirmed the recently proposed picture of ions mutually
interlocking between the rods.  The closest ions along the direction
of the rod have a tendency to be condensed on different rods, even
though defects in this structure occur quite frequently.  In agreement
with earlier observations on discretized models \cite{DiCa01} this
structure is remarkably short ranged.  While attractive interactions
have been found for all cases in which a first correlation peak is
discernible, the mere existence of a first correlation hole was
demonstrated to be insufficient.

The phenomena observed in our molecular dynamics simulations also give
rise to several new questions, for instance: How can one predict the
relative contribution of electrostatic and short range forces to the
net interaction between rods? The structural changes in the azimuthal
correlation functions result from usual pair correlation functions
being ``wound up'' around the rod.  In consequence, the condensed ions
predominantly push or pull from specific directions, not necessarily
aligned with the direction joining the rods.  What effects on the net
force does this have? Under which circumstances do the ions between
the rods get pulled away from the surface, and when is this the
dominant force of attraction?  Further studies along these lines are
currently under way.


\section{Acknowledgments}

We would like to thank B.\ J\"onsson and B.\ Shklovskii for
stimulating discussions. MD thanks the German Science Foundation (DFG)
for financial support under grant De$\,$775/1-1. CH wants to thank the
``Zentrum f\"ur Mul\-ti\-funk\-tio\-nel\-le Werk\-stoffe und
Mi\-ni\-atu\-ri\-sier\-te Fun\-ktions\-ein\-hei\-ten", grant BMBF 03N
6500.




\end{document}